%% file: genbem_iop.tex
\documentclass[12pt]{iopart}
\usepackage{graphicx}  
\usepackage{harvard}  
\usepackage{gensymb,xcolor}
\input{genbem_commanddefs}
\begin{document}
\title[BEM for general piece-wise homogeneous volume conductor]{Integral equations and boundary-element solution for static potential in a general piece-wise homogeneous volume conductor}
\author{Matti Stenroos}
\address{Department of Neuroscience and Biomedical Engineering, Aalto University, P.O.~Box~12200, FI-00076 Aalto}
\ead{matti.stenroos@aalto.fi}
\vspace{10pt}
\begin{indented}
\item[] 1st September 2016
\end{indented}

\begin{abstract}
Boundary element methods (BEM) are used for forward computation of bioelectromagnetic fields in multi-compartment volume conductor models. Most BEM approaches assume that each compartment is in contact with at most one external compartment. In this work, I present a general surface integral equation and BEM discretization that remove this limitation and allow BEM modeling of general piecewise-homogeneous medium. The new integral equation allows positioning of field points at junctioned boundary of more than two compartments, enabling the use of linear collocation BEM in such a complex geometry. A modular BEM implementation is presented for linear collocation and Galerkin approaches, starting from standard formulation. The approach and resulting solver are verified in
\textcolor{black}{ four ways, including comparisons of volume and surface potentials to those obtained using the finite element method (FEM)},
and the effect of a hole in skull on electroencephalographic scalp potentials is demonstrated. 
\end{abstract}
\vspace{2pc}
\noindent{\it Keywords}: boundary element method, bioelectromagnetism, forward problem,
electrostatics, numerical field computation, volume conductor model
%\submitto{\PMB}
%\maketitle
\section{Introduction}
\begin{figure}[!tb]
\centerline{\includegraphics{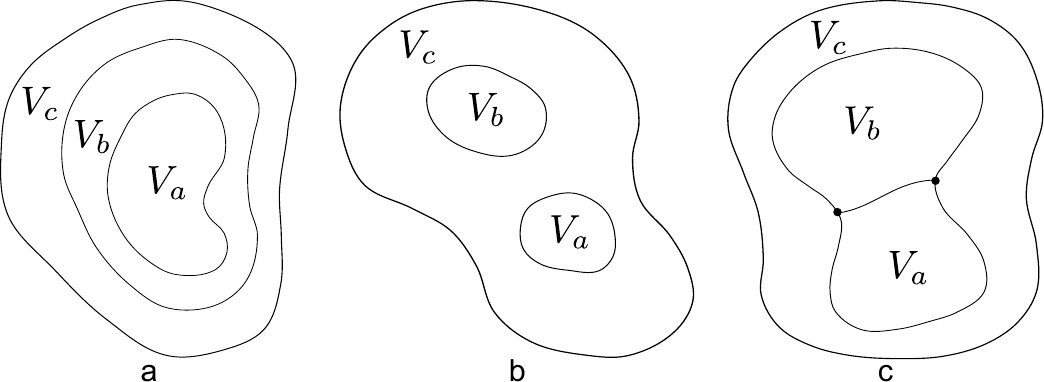}}
\caption{Piecewise homogeneous models that have three compartments $V_a, V_b$, and $V_c$: a) nested, b) non-nested, c) junctioned. The black dots mark the junctions.}
\label{geotypes}
\end{figure}

Boundary element methods (BEM) are used for forward modeling of bioelectric and biomagnetic fields, when piecewise homogeneous medium can be assumed. In most BEM formulations, there is, however, a major restriction: the change of conductivity across any closed boundary of a homogeneous compartment needs to be constant. Examples of such boundary geometries are shown in figure \ref{geotypes}a and \ref{geotypes}b. Figure \ref{geotypes}c shows a case where the jumps of conductivity across compartment boundaries are not constant; boundaries contain junctions, and standard BEM approaches are not applicable. Junctioned geometry is needed for realistic modeling of, {\it e.g.}, holes in the skull, fontanels and sutures in infant head, boundary of ventricular myocardium, blood and thorax, and of epicardial fat pads. So far this kind of geometries have been modeled either further simplified \cite{Oostenveld02} or using volume-based methods \cite{Lau16,Lew13,Keller10}. In this note, I present a general surface integral equation and BEM discretization that can be used with any piecewise homogeneous geometry. 

Integral equations that enable the BEM solution of bioelectromagnetic problems were derived in the 1960s without explicitely considering junctioned geometries \cite{Barr66,Barnard67a}.
The conceptual basis for these works was laid in \cite{Gelernter64}, where also the first discretized thorax model was presented.
An early boundary-element approach was formulated in \cite{Barnard67b}; even though the constant-potential formulation was done for non-junctioned geometry, that  approach is actually compatible also with junctioned geometry.
Junctioned geometry was briefly treated in \cite {Hamalainen93}, but the field point was assumed to lie on a smooth region of the surface. To use linear or higher-order basis functions and computationally efficient collocation BEM \cite{deMunck92,Stenroos07}, the integral equations, however, need to be evaluated at vertices of the meshed boundary, including the junction vertices, for which the smooth formulation does not apply.
\textcolor{black}{In addition, the junction vertices need a special treatment when assembling the BEM matrix}.

BEM computations in junctioned geometry were shown in \cite{Akalin04}, but all detail about the theory and implementation was omitted. In \cite{Kybic06}, 
\textcolor{black}{
integral equations of the symmetric BEM approach were formulated for a non-nested model, but the treatment of junctions in the assembly of the transfer matrix was omitted; the method description thus does not cover junctioned models, even though results for a junctioned example case were shown. In addition,} BEM matrix composition for a general geometry was not presented.  
This note presents the integral equations and matrix composition for junctioned BEM in a general form.

\section{Methods}
Consider quasi-static source current density $\Js$ in a resistive medium of conductivity $\sigma$. The sources and sinks of $\Js$ are associated with charge density that gives rise to electric field $\vec E=-\grad\phi$, where $\phi$ is the electric potential. The electric field drives volume current density $\Jv$, leading to a total current of $\vec J=\Js+\Jv=\Js-\sigma\grad\phi$. Applying the law of charge conservation, we get the Poisson equation $\div(\sigma\grad\phi)=\div\Js$.
In a piece-wise homogeneous medium, we have for each compartment
\beq
\nabla^2\phi=-\iv/\sigma,
\label{poisson}
\eeq
where $\iv=-\div\Js$ is the source density of volume current. At boundary $S$, the potential and normal component of $\Jv$ are continuous,
\numparts
\beqa
\label{bcV}
\phi(\vr\rightarrow S_-)=\phi(\vr\rightarrow S_+)\\
\label{bcGamma}
\sigma_-\frac{\dee\phi(\vr\rightarrow S_-)}{\dee n}=\sigma_+\frac{\dee\phi(\vr\rightarrow S_+)}{\dee n},
\eeqa
\endnumparts
where derivative is taken in the direction of outer normal $\bi n$, subscripts $\pm$ label regions outside and inside $S$, and $\vr\rightarrow S_\pm$ means that $\vr$ approaches $S$ from outside and inside, respectively. 
Define $\Gamma^S_\pm= \dee\phi(\vr\rightarrow S_\pm)/\dee n$.
\subsection{Integral equations}

Study a compartment $V_i$ of conductivity $\sigma_i$ bounded by closed surface $\partial V_i$. Label compartments and surfaces using sub- and superscripts, respectively.
Using Green's theorem and (\ref{poisson}), we get \cite{Stenroos09}
\beq
\Omega^{\dee V_i}(\vr)\phi(\vr)=v_i(\vr)+(S^{\dee V_i}\Gamma^{\dee V_i}_-)(\vr)-(D^{\dee V_i}\phi)(\vr),
\label{base}
\eeq
where $\Omega^{\dee V_i}(\vr)$ is the normalized solid angle spanned by the boundary of $V_i$ at $\vr$,
$v_i$ is the potential produced in an infinite homogeneous medium of conductivity $\sigma_i$ by sources within $V_i$, 
$S^i$ is the single-layer operator,
and $D^j$ is the double-layer operator:
\beqa
v_i(\vr)=\frac{1}{4\pi\sigma_i}\int\limits_{V_i}\frac{\iv(\vr\pr)}{\absR}\dV\pr\\
(D^jf)(\vr)=\frac{1}{4\pi}\int\limits_{S^j}f(\vr\pr)\frac{(\R)}{\absR^3}\cdot\vdS\pr\\
(S^jf)(\vr)=\frac{1}{4\pi}\int\limits_{S^j}\frac{f(\vr\pr)}{\absR}\dS\pr.
\eeqa
\begin{figure}[!tb]
\centerline{\includegraphics{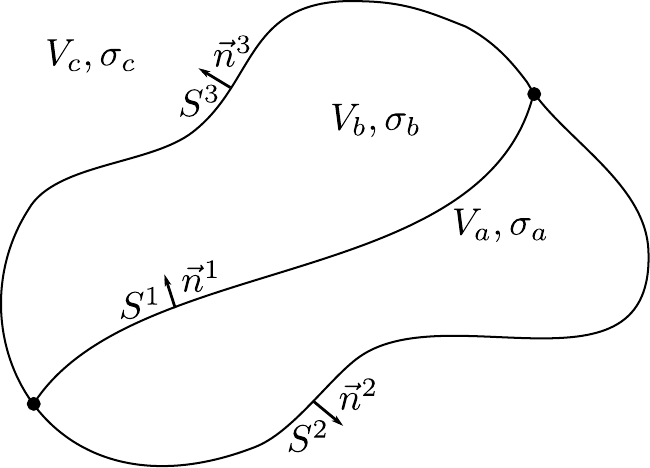}}
\caption{A volume conductor with junctioned boundaries. The compartment boundaries are expressed using open surfaces $S^i$, and the black dots mark the junctions, where all three boundary surfaces join.}
\label{geo}
\end{figure}

Now consider a two-compartment ($V_a, V_b$) volume conductor immersed in infinite homogeneous space $V_c$ as illustrated figure \ref{geo}. Compartments $a$ and $b$ are connected with each other via surface $S^1$ and with compartment $c$ via surfaces $S^2$ and $S^3$, respectively. All surfaces are open and join each other at junctions, which form the boundaries of the surfaces. Applying (\ref{base}) to compartments $a,b$ and $c$, we get 
\numparts
\beqa
\label{a1}
\Omega^{\dee V_a}\phi(\vr)\!=\!v_a(\vr)\!+\!(S^1\Gamma^1_-)(\vr)\!+\!(S^2\Gamma^2_-)(\vr)\!-\!
(D^1\phi^1)(\vr)\!-\!(D^2\phi^2)(\vr)\\
\label{b1}
\Omega^{\dee V_b}\phi(\vr)\!=\!v_b(\vr)\!-\!(S^1\Gamma^1_+)(\vr)\!+\!(S^3\Gamma^3_-)(\vr)\!+\!
(D^1\phi^1)(\vr)\!-\!(D^3\phi^3)(\vr)\\
\Omega^{\dee V_c}\phi(\vr)\!=\!v_c(\vr)\!-\!(S^2\Gamma^2_+)(\vr)\!-\!(S^3\Gamma^3_+)(\vr)\!+\!
(D^2\phi^2)(\vr)\!+\!(D^3\phi^3)(\vr)
\eeqa
\endnumparts
Next, multiply each equation by the conductivity of the corresponding compartment.
For surface terms, express the conductivity in terms of corresponding surface, labeling compartments inside and outside the surface $k$ as $\sigma^k_-$ and $\sigma^k_+$, respectively; for example, $\sigma_b=\sigma^1_+=\sigma^3_-$. Sum the equations and use (\ref{bcGamma}) to get
\beq
\sum\limits_{i=a}^c\sigma_i\Omega^{\dee V_i}\phi(\vr)=
\sum\limits_{i=a}^c \sigma_i v_i(\vr)+
\sum\limits_{l=1}^3 (\sigma^l_+-\sigma^l_-)(D^l\phi^l)(\vr).
\label{genintexample}
\eeq
The first right-side term results in a conductivity-independent potential function; we express it using infinite-medium potential $\phi_\infty$ in a medium that has dummy (unit) conductivity $\sigma_\mathrm{s}$. Equation (\ref{genintexample}) generalizes directly to $M$ compartments and $N$ boundary surfaces, resulting in the general form of surface integral equation for scalar potential:
\beq
\sum\limits_{i=1}^M\sigma_i\Omega^{\dee V_i}\phi(\vr)=
\ss\phi_\infty(\vr)+\frac{1}{4\pi}
\sum\limits_{l=1}^N 
(\sigma^l_+-\sigma^l_-)\int\limits_{S^l}\phi(\vr\pr)\frac{(\R)}{\absR^3}\cdot\vdS\pr,
\label{genint}
\eeq
where
\beq
\phi_\infty(\vr)=\frac{1}{4\pi\ss}\int\limits_{\mathrm{all\ }\iv}\frac{\iv(\vr\pr)}{\absR}\dV\pr
=
\frac{1}{4\pi\ss}\int\limits_{\mathrm{all\ }\Js}\frac{\Js(\vr\pr)\cdot(\R)}{\absR}\dV\pr.
\label{phiinf}
\eeq
Evaluating the left side of (\ref{genint}) with different positionings of $\vr$ but avoiding junctions, the equation reduces to previously presented equations:
\begin{itemize}
\item
For $\vr\in V_j,\vr\notin\dee V_j$, we have $\Omega^{\dee V_j}(\vr)=1$ and $\Omega^{\dee V_k}(\vr)=0$, $j\neq k$, yielding the equation presented by \citename{Geselowitz67} \citeyear{Geselowitz67}.
\item
For $\vr\in S^k$ and $\vr\notin S^j, j\neq k$, the left side yields
$(\sigma^k_+\Omega^{\dee V_+^k}+\sigma^k_-\Omega^{\dee V_-^k})\phi(\vr)$,
where $V^k_\pm$ label compartments outside and inside $S^k$, respectively. 
The result matches the sharp-edged form \cite{Ferguson97,Stenroos09}.
\item
If $\vr\in S^k$ and $\vr\notin S^j, j\neq k$, and $S^k$ is smooth around $\vr$, we get the form presented by \citename{Barnard67a} \citeyear{Barnard67a} and in junctioned geometry by \cite{Hamalainen93},
\beq
\frac{1}{2}(\sigma^k_++\sigma^k_-)\phi(\vr)\!=\!
\ss\phi_\infty(\vr)+
\frac{1}{4\pi}\!\sum\limits_{l=1}^N (\sigma^l_+-\sigma^l_-)\!
\int\limits_{S^l}\phi(\vr\pr)\frac{(\R)}{\absR^3}\cdot\vdS\pr.
\label{surfint}
\eeq
\end{itemize}
\subsection{Boundary-element discretization}
\label{bemtext}
A boundary-element model is built by tessellating boundary surfaces into polygon meshes,
approximating potential $\phi$ on all boundary surfaces as a linear combination of a set of basis functions defined around a total of $N_\mathrm d$ discretization points, evaluating (\ref{genint}) on all boundary surfaces, and minimizing the residual of the approximated potential with respect to $N_\mathrm d$ weight functions. This results in an equation of the form
$\T\p=\pinf$,
where $(N_\mathrm d\times 1)$-vectors $\p$ and $\pinf$ contain the values of $\phi$ and weighted $\sigma_\mathrm s\phi_\infty$ in all discretization points, and $\T$ is a $(N_\mathrm d\times N_\mathrm d)$ matrix. The process is thoroughly explained in, {\it e.g.}, \cite{Stenroos07,Stenroos09}, and the assembly of $\T$ and $\pinf$ 
as used in this work is shown compactly in the Appendix.

In this work, I use triangle meshes, linear basis functions, and collocation and Galerkin weightings \cite{deMunck92,Mosher99}: Triangle mesh $k$ that contains $N^k_\mathrm v$ vertices and $N^k_\mathrm t$ triangles is defined by an $(N^k_\mathrm v \times 3)$ array of vertex coordinates and an $(N^k_\mathrm t \times 3)$ array of indices that point to rows of the vertex array ("local indexing"), and a linear basis function is spanned on all triangles that belong to one vertex $\bi v^k_i$ so that the function gets value 1 in the target vertex and decreases linearly to value 0 in the neighboring vertices.
$N_\mathrm d$ is thus equal to the total number of vertices $\sum_{k=1}^N N^k_\mathrm v$ in the model. In collocation weighting, Dirac delta functions defined in vertices are used as weight functions, while in Galerkin weighting, the same functions are used as basis and weight functions \cite{Mosher99,Stenroos08}.
In $\p,\pinf$ and $\T$, vertices of each mesh are concatenated together so that the vertices of $k^\mathrm{th}$ mesh have pooled indices of
$[(\sum_{i=1}^{(k-1)} N^i_\mathrm v)+1,\sum_{i=1}^k N^i_\mathrm v]$.

In non-junctioned geometry, meshes are closed and fully separate; every row of $\p$ and $\pinf$ refers to a unique vertex, and every element of $\T$ refers to a unique pair of two vertices.
In junctioned geometry, the meshes that are connected to a junction are open, and vertices and element edges at junctions are shared by at least two meshes. Using standard BEM implementation with these meshes, many rows of $\p$ and $\pinf$ and many elements of $\T$ thus refer to the same vertices or vertex pairs. Now consider a junctioned geometry. The standard BEM matrix is of the form
\beq
\Tl\phil=\pinfl,
\label{bemlocal}
\eeq
where subscript p refers to pooled indexing. Collect all unique vertices to a global set of $N_\mathrm u$ vertices, $N_\mathrm u<N_\mathrm d$.
Then define matrix operators that convert from pooled to global indexing in either summing or extracting way: operator $\Ilg$ ($N_\mathrm u \times N_\mathrm d$ matrix) converts data in rows from pooled to global indexing and sums the contributions from junction vertices, while $\Elg$ does the same but extracts the contribution from only one of the junction vertices. Operator $\Igl$ converts from global to pooled indexing, setting the global values to all corresponding vertices. It turns out that $\Igl=\Ilg^\mathrm T$, where $^\mathrm T$ marks transposis.

In linear collocation (LC) BEM, each row of (\ref{bemlocal}) refers to one vertex. The rows that map to the same global vertex contain the same information, and we can use the $\Elg$ operator to convert to global indexing:
\beq
\Elg\Tl\phil=\Elg\pinfl=\pinfg.
\eeq
Each column of $\Tl$ models the contribution of the neighborhood of a vertex. For junction vertices, these neighborhoods are different for different meshes. To convert $\Tl$ to operate on global indexing, we sum the contributions of each local vertex using $\Ilg$. To apply $\Ilg$ to columns, transpose it and multiply $\Tl$ from the right side,
\beq
\Elg\Tl\Ilg^\mathrm T\phig=\pinfg.
\eeq
Another way to explain this step is to just say that we express $\phil$ as $\Igl\phig=\Ilg^\mathrm T\phig$. We now get the final transfer matrix $\Fg$ in global indexing:
\beq
\phig=\underbrace{(\Elg\Tl\Igl)^{-1}}_{\Fg}\pinfg.
\label{TgLC}
\eeq

For linear Galerkin (LG) BEM, the logic is the same, but as also rows characterize neighborhoods, we need to use the summing index conversion:
\beqa
\Ilg\Tl\Ilg^\mathrm T\phig=\Ilg\pinfl\\
\Rightarrow\phig=\underbrace{(\Ilg\Tl\Ilg^\mathrm T)^{-1}\Ilg}_{\F_{\mathrm{\textcolor{black}{p2g}}}}\pinfl.
\label{TgLG}
\eeqa
Here $\pinfl$ is kept because it can be directly implemented using standard BEM routines and local meshes, while implementing the LG version of $\pinfg$ would demand the construction of a junctioned global mesh, adding unnecessary complexity.

\section{Results}
\subsection{\textcolor{black}{Homogeneous sphere}}
I implemented the index conversion operators and formulas (\ref{TgLC}) and (\ref{TgLG}) in Matlab (version R2014b, www.mathworks.com) using existing BEM tools \cite{Stenroos07,Stenroos08,Stenroos09} and carried out three types of verifications:

First, I verified the index conversions by randomly splitting the boundary of a spherical single-compartment model to three open meshes, building BEM models using both the original closed mesh (model 1) and the separate open meshes (model 2), and simulating a set of random dipolar test sources; the resulting surface potentials in models 1 and 2 were identical up to numerical precision (relative difference RDIF\footnote
{RDIF$= |\bd_\mathrm{test}-\bd_\mathrm{ref}/|\bd_\mathrm{ref}|$, where $\bd_\mathrm{test}$ and $\bd_\mathrm{ref}$ are test and reference solutions in vector form.}
$\sim 10^{-15}$).
Next, I split a spherical model to two open half-spheres at xy plane and added a boundary between the halfs, resulting in a two-compartment three-mesh model as illustrated in figure \ref{results}. First, I set $\sigma_1=\sigma_2$ (model 3), resulting in a model physically identical to model 1, 
and evaluated the outer-boundary potential of a random set of test sources. 
With the LC approach, the results of model 3 should be nearly identical to model 1, and indeed they were (RDIF~$\sim 10^{-15}$). With LG approach, differences are expected to occur near the junctioned boundary at $z=0$, because the potential for the junction vertices is integrated in different neighborhoods in models 1 and model 3. These differences were, however, much smaller than overall errors of these models as compared to the closed-form analytical formula.
\subsection{\textcolor{black}{Volume potential in split sphere}}
\begin{figure}[!tb]
\centerline{\includegraphics[width=14 cm]{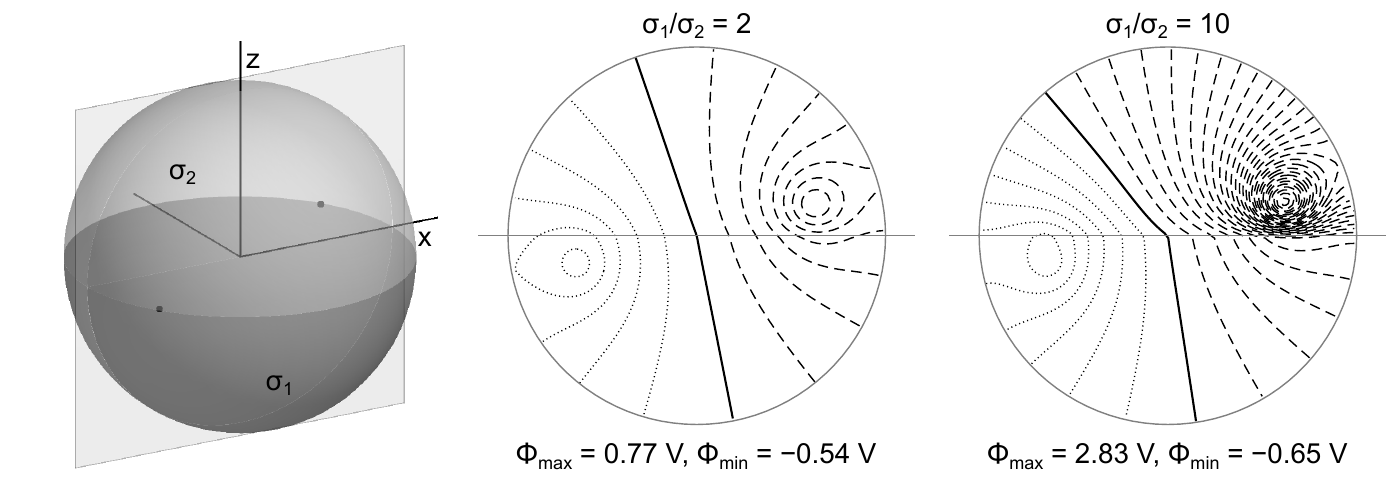}}
\caption{Geometry of the split-sphere model (left) and example results with two different conductivities in the upper half-sphere (middle, right). Contour step is 0.1 V, and the upper half is at lower conductivity.}
\label{results}
\end{figure}

To verify the weighting of conductivity terms, I set $\sigma_1$ to 1~S/m and $\sigma_2$ to 0.5~S/m or 0.1~S/m, and implemented the same models using Comsol Multiphysics software (version 5.1, www.comsol.com) that uses finite element method (FEM). The simulation scenario had to be easy for the FEM to handle and clearly show the effect of the conductivity boundary. To achieve this, I placed a monopolar sink and source symmetrically around the origin, rather close to the split but far from each other (source locations $\pm(8, 8, 2)$ cm, strengths  $\pm0.1$ A, sphere radius 13 cm), and evaluated potential on the xz plane; the BEM solution for the volume was obtained by first solving potential on all boundary surfaces and then using (\ref{genint}) with field points set in the evaluation plane. The results are shown in figure \ref{results}. The field patterns obtained using the highest-resolution physics-controlled mesh in Comsol (''extremely fine'', degrees of freedom $\mathrm{DoF}=1542049$) and BEM ($\mathrm{DoF}=3205$) were nearly identical
($0.001<\mathrm{RDIF}<0.003$), suggesting that the BEM solver is correctly implemented. The difference between Comsol ''extremely fine'' and ''fine'' ($\mathrm{DoF}=23061$) meshings was
slightly larger ($0.009<\mathrm{RDIF}<0.018$).
% new text begins here
\subsection{\textcolor{black}{Spherical head model: hole in skull}}
\begin{figure}[!tb]
\centerline{\includegraphics[width=12 cm]{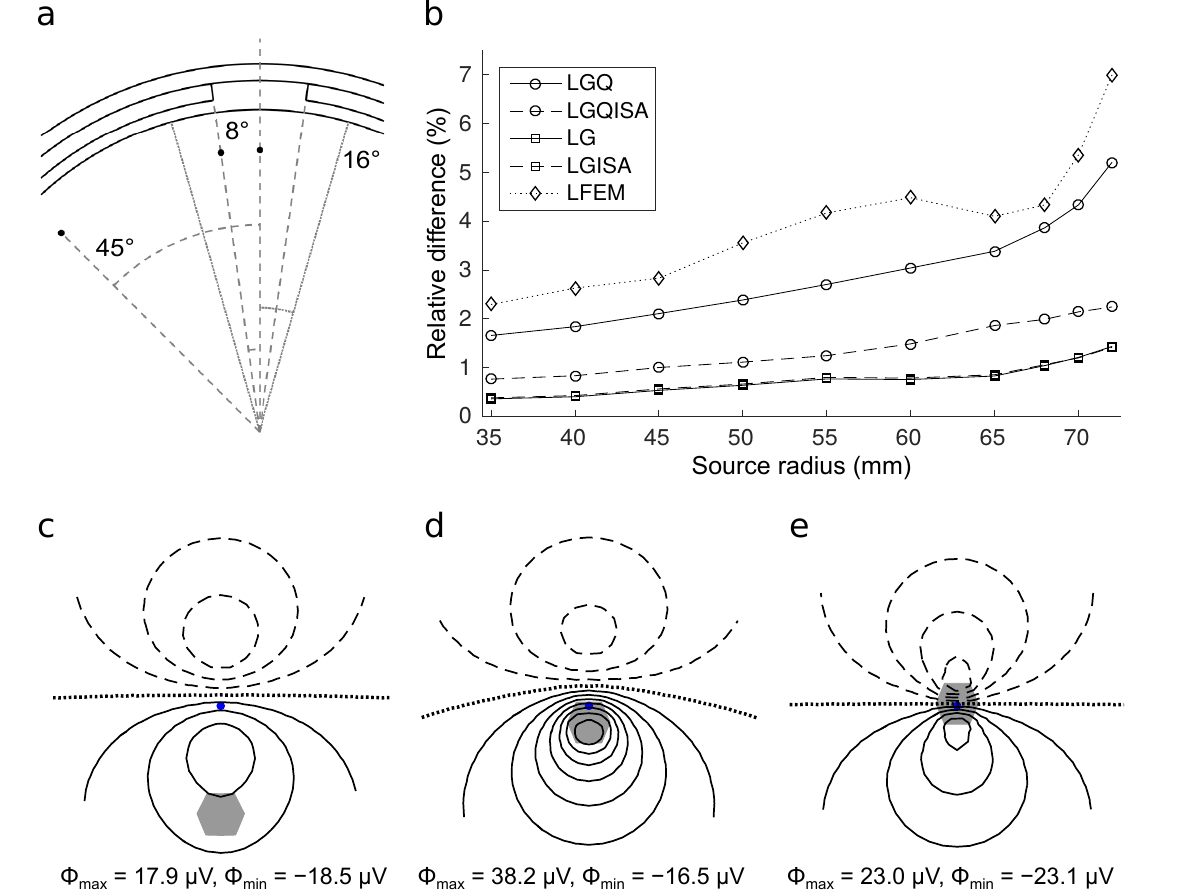}}
\caption{Effect of a hole in the skull. Plot a) shows the model geometry; example sources are marked with black points \textcolor{black}{and the space in which the random source positions are is delineated with dotted lines. Plot b) shows the median of RDIF as function of source depth, using QFEM as the reference solution.} Plots c)--e) show scalp topographies of a tangential dipole, when the dipole is c) quite far from the hole at 45\degree, d) under the boundary of the hole 8\degree, and e) under the center of the hole. The topographies are viewed from above the source, marked with a gray point, and the location of the hole is drawn in light gray. Maxima and minima of each topography are marked below the plots, and the contour step is 5 $\mathrm{\mu V}$.}
\label{example}
\end{figure}

Consider a four-shell (brain, cerebrospinal fluid CSF, skull, scalp) volume conductor, where the skull has a small hole that is filled with CSF.
\textcolor{black}{In this simulation, I compare surface potentials obtained using FEM and four Galerkin-based BEM approaches}.
I built a spherical concentric four-shell model (radii 78, 81, 85, and 89 mm, conductivities 0.33, 1.79, 0.0132, and 0.33 S/m), manually made a cavity (mean diameter 23 mm, opening angle $\approx 16\degree$) through the skull compartment,
\textcolor{black}
{and included the cavity in the CSF compartment. The model geometry is illustrated figure \ref{example}a.
I implemented the model in Comsol using default settings for meshing and solver (normal physics-controlled meshing, quadratic solver (QFEM, $\mathrm{DoF}=102802$) and also a linear solver (LFEM, $\mathrm{DoF}=13470$) and extracted the boundary meshes for the BEM models (number of vertices $=\mathrm{DoF}=9801$). The LFEM had the same surface elements and approximation order as the Galerkin BEM and thus enables direct and fair comparison of surface and volume approaches. Using a small set of test dipoles, I assessed the error of the QFEM as function of source depth against a denser FEM model ($\mathrm{DoF}=240329$) and considered QFEM an adequate reference model, when sources are at max. radius of 72 mm, where the median RDIF between QFEM and the denser FEM was 1\%.}

\textcolor{black}{
I generated a random set of 100 source positions on a spherical cap that was centered around the cavity and had a radius of 16\degree; approximately one third of the sources were under the cavity. For these template sources, I assigned randomly-oriented unit-strength dipole moments and projected the sources at 10 different depths, totaling in 1000 test sources. I solved the surface potentials of all sources using the QFEM, LFEM and BEMs built with four different linear approaches: The standard Galerkin (LG) with 13 quadrature points per triangle, the simpler and faster Quick Galerkin (LGQ) \cite{Stenroos12a} that has only one quadrature point per triangle, and isolated-source formulations of these (LGISA, LGQISA) that used approach 3 of \cite{Stenroos12b}. Then, I computed relative differences between the QFEM model and the other models. The results as function of source depth are shown in figure \ref{example}b. LG and LGISA perform identically well, with median RDIF below 1.2\% for all source depths. LGQISA stays below 1\% at depths below 60 mm; with more superficial sources the error increases to about 2.2\%. LGQ without ISA is clearly less accurate, when sources approach the pial boundary and skull. LFEM is the least accurate of the tested approaches. The maximum RDIFs of BEMs and LFEM were approximately 1.5--2.2 and 2.8 times the median RDIF, respectively.
}

\textcolor{black}{
Finally, I illustrate the effect of such a hole on electroencephalographic scalp potentials:
I placed tangential dipoles (dipole moment 100 nAm) at the depth of 10 mm from the CSF boundary at three positions 1) far from the hole at 45\degree angle from the center of the hole; 2) under the hole boundary at 8\degree, and 3) under the hole center at 0\degree. The scalp potentials are shown in figure \ref{example}c--e. For the source at 45\degree (c), the scalp topography is highly similar to that produced without the hole. When the source is under the boundary of the hole (d), the hole strongly affects the shape of the topography and the overall effect of the hole is at largest.}

% new text ends here
\section{Discussion}
In this work, I have derived a surface integral equation for static potential in a general piece-wise homogeneous conductor model, described the boundary-element discretization of this equation using both linear collocation (LC) and linear Galerkin (LG) approaches, and implemented and verified the approach. To my knowledge, this is the first work to 1) present the integral equation for potential in a junctioned geometry in a form that allows linear collocation BEM solution,
\textcolor{black}
{to 2) fully describe, how non-unique vertices are treated in BEM matrix construction and inversion}, and to 3) present BEM matrix construction for general piece-wise homogeneous geometry. With this general BEM (gBEM) formulation, any topological construction of boundaries, including meshes with disjoint triangle sets, is allowed as long as the junctioned or shared boundaries are expressed using open non-overlapping meshes that connect to each other via their shared vertices and triangle edges.

In addition to volume conductor studies, this gBEM approach can be used with any phenomena that obey Poisson's equation (\ref{poisson}) and have boundary conditions of the form (\ref{bcV}--\ref{bcGamma}). For example, to solve electrostatic problems in dielectric medium, just replace $\sigma$ with permittivity $\epsilon$ and $i_\mathrm{v}$ with free charge $\rho_\mathrm{f}$. As the BEM solution is computed by multiplying the infinite-medium potential with the BEM matrix, the approach and solver also suits for solving problems, in which the source field is directly expressed in terms of potential in infinite medium (for example, a scatterer placed in initially homogeneous electric field).

The basic BEM equations presented in the Appendix are in slightly different form than in earlier works, but functionally they are identical to those given in \cite{Stenroos09}. In this work, the geometry and conductivity parameters are fully decoupled and each conductivity term depends on one surface index only, and the sharp-edged LC BEM is presented in a simpler form. Depending on computational platform, it may, however, be more efficient to implement the formulas using row/columnwise multiplications instead of the diagonal $\mathbf \Sigma$ matrices.
Once $\phig$ is computed and converted to pooled local indexing, magnetic field can be computed using open meshes and integral equation for magnetic field as detailed in \cite{Ferguson94} and implemented in Helsinki BEM library \cite{Stenroos07}.

\textcolor{black}{
The verification results presented in figure \ref{example}b match well the verification presented in \cite{Stenroos12a,Stenroos12b}, only with slightly smaller overall error that is likely due to numerically easier boundary value problem (the hole in the skull provides a high-conductivity pathway to the scalp, while in intact skull, all currents have to pass through the low-conductivity skull). When the same approximation order and discretization of boundary surfaces was used for both the FEM and BEM, the BEM performed better, even though the meshing was optimized for the FEM. This was expected: the BEM discretizes the governing equation only at conductivity boundaries, while the FEM discretizes the equation in the whole volume, including the source space. The differences between the approaches are, however, small, compared to the model errors due to simplifications and erroneous conductivities; see \cite{Stenroos16}.
}

The modifications that enable the use of junctioned boundaries in a standard LC or LG BEM solver are straightforward to do using index-conversion operators that are applied before and after inverting the system matrix $\T$; the element integrals or assembly of $\T$ do not need to be changed. Meshing may, however, be more complicated than in traditional BEM geometries, as many surface meshing tools cannot cope with junctioned geometry. If no suitable surface mesher is available, the meshing can be done in two steps, first using a volume-based finite-element mesher and then extracting the boundaries, \textcolor{black}{like was done in this study}. Making controlled surface meshes using a FEM mesher requires, however, expertise. One surface-based approach to meshing junctioned boundaries was presented in \cite{Akalin04}, but those algorithms are, to my knowledge, not publicly available. I leave further study of meshing for future research.

To advance developing and evaluation of surface meshing tools and understanding and use of BEM methodology, the LC \textcolor{black}{and LGQ gBEM solvers, including LGQISA,} are available from the author for academic use.

\ack
I thank Mr. Lari Koponen for discussions and for helping with Comsol software and Academy of Finland (290018) for partial funding of this study.
\appendix
\section{Composition of BEM matrices}
Discretization of Eq. (\ref{surfint}) using linear Galerkin (LG) approach as in \cite{Stenroos08,Stenroos09} results in
\def\bbm{\left[\begin{array}}
\def\ebm{\end{array}\right]}

\beq
\eqalign{
\fl
%\begin{array}{ll}
\overbrace{
\bbm{ccc}
	 \Sa^1\\
	& \ddots\\
	&& \Sa^N
\ebm}^{\Sa}
\overbrace{
\bbm{ccc}
	 \bA^1\\
	& \ddots\\
	&& \bA^N\\
\ebm}^{\bA}
\overbrace{
\bbm{c}
\bp^1\\\vdots\\\bp^N\\
\ebm}^\bp
=\cr
\overbrace{
\bbm{c}
\bp^1_\infty\\\vdots\\\bp^N_\infty\\
\ebm}^{\bp_\infty}+
\overbrace{
\bbm{ccc}
\bD^{11}&\cdots&\bD^{1N}\\
\vdots&\ddots&\vdots\\
\bD^{N1}&\cdots&\bD^{NN}\\
\ebm}^\bD
\overbrace{
\bbm{ccc}
	 \Sd^1\\
	& \ddots\\
	&& \Sd^N\\
\ebm}^{\Sd}
\bbm{c}
\bp^1\\\vdots\\\bp^N
\ebm,
}
%\end{array}
\eeq
where $\bp^k_i=\phi(\bi v^k_i)$,
$
\Sa^k(i,j)=\frac{1}{2}(\sigma^k_++\sigma^k_-)\delta_{ij}
$,
$
\Sd^k(i,j)=(\sigma^k_+-\sigma^k_-)\delta_{ij},
$
and other elements are described in Table \ref{operators}.
Thus, we get
\beq
\label{TLG}
\overbrace{
(\Sa\bA-\bD\Sd)}
^{\T_\mathrm{LG}}\bp=\bp_\infty.
\eeq
For linear collocation (LC) BEM, discretize (\ref{genint}) following \cite{Stenroos07,Stenroos09} to get
\beq\label{TLC}
\bS_\Omega\bp
=\bp_\infty+\bD\Sd\bp
\Rightarrow
\overbrace{
(\bS_\Omega-\bD\Sd)}
^{\T_{\mathrm{LC}}}\bp=\bp_\infty.
\eeq
where $\bp, \bp_\infty,\bD$, and $\Sd$ have the same structure as above, and
\beq
\bS_\Omega=
\sum\limits_{j=1}^M
\sigma_l
\bbm{ccc}
	\bO^1_j\\
&\ddots\\
&&\bO^N_j
\ebm.
\eeq
Then, after setting the level of zero potential \cite{Fischer02}, we can invert the $\T$ matrix and solve the unknown potential,
\beq
\bp=\overbrace{\T^{-1}}^\F\bp_\infty.
\eeq
\begin{table}[!h]
\caption{Elements of BEM matrices and vectors}
\label{operators}
\vspace{4mm}
\begin{tabular}{c|c|c|c|c}
	 & $\bA^k(i,j)$ & $\bO^k_l(i,j)$ & $\bp^k_\infty(i)$ & $\bD^{kl}(i,j)$\\
\hline
LG\phantom{$\int\limits^b$} & $\int_{S^k}\psi^k_i\psi^k_j$\dS & &$\sigma_s\int_{S^k}\psi^k_i\phi_\infty\dS$ & $\int_{S^k}\psi^k_i (D^l\psi^l_j)\dS$
\\
LC\phantom{$\int\limits^b$} & & $\Omega_{\partial V_l}(\vec v^k_i)\delta_{ij}$& $\sigma_s\phi_\infty(\vec v^k_i)$ &$(D^l\psi^l_j)(\vec v^k_i)$\\
\end{tabular}
\end{table}
\section*{References}
\bibliography{genbem_refs}
\end{document}

%% file: genbem_commanddefs.tex
\def\div{\nabla\cdot}
\def\grad{\nabla}
\def\dee{\partial}

\def\iv{i_\mathrm{v}}
\def\Js{\vec J_\mathrm{s}}
\def\Jv{\vec J_\mathrm{v}}

\def\ss{\sigma_\mathrm{s}}

\def\vr{\bi r}

\def\R{\vr-\vr\pr}
\def\absR{\vert\vr-\vr\pr\vert}

\def\pr{^{\,\prime}}

\def\dV{\,dV}
\def\dS{\,dS}
\def\vdS{\,\vec dS}

\def\beq{\begin{equation}}
\def\eeq{\end{equation}}
\def\beqa{\begin{eqnarray}}
\def\eeqa{\end{eqnarray}}

\def\T{\mathbf{T}}

\def\F{\mathbf{F}}
\def\Fg{\mathbf{F}_\mathrm{g}}

\def\p{\mathbf{\Phi}}
\def\pinf{\mathbf{\Phi}^\infty}

\def\phig{\mathbf{\Phi}_\mathrm{g}}
\def\pinfg{\mathbf{\Phi}^\infty_\mathrm{g}}

\def\phil{\mathbf{\Phi}_\mathrm{p}}
\def\pinfl{\mathbf{\Phi}^\infty_\mathrm{p}}
\def\Ilg{\mathbf{I}_\mathrm{p2g}}
\def\Elg{\mathbf{E}_\mathrm{p2g}}
\def\Igl{\mathbf{I}_\mathrm{g2p}}

\def\Tl{\mathbf{T}_\mathrm{p}}

\def\Sa{\mathbf{\Sigma}_\mathrm{ave}}
\def\Sd{\mathbf{\Sigma}_\Delta}
\def\bp{\mathbf{\Phi}}
\def\bD{\mathbf{D}}
\def\bA{\mathbf{A}}

\def\bO{\mathbf{\Omega}}

\def\bS{\mathbf{\Sigma}}
\def\bd{\mathbf{d}}